\documentclass[review]{elsarticle}

\usepackage{lineno,hyperref}
\modulolinenumbers[5]

\journal{Journal of Physics and Chemistry of Solids}

\begin{document}

\begin{frontmatter}

\title{Ground state properties of a two dimensional Fermi superfluid with an anisotropic spin-orbit coupling }

\author{Kezhao Zhou\corref{cor1}}\ead{kezhaozhou@gmail.com}
\author{Zhidong Zhang\corref{cor1}}\ead{zdzhang@imr.ac.cn}

\address{Shenyang National Laboratory for Materials Science, Institute of Metal
Research, Chinese Academy of
Sciences, 72 Wenhua Road, Shenyang 110016, China}

\cortext[cor1]{Corresponding author.}

\begin{abstract}
We performed a theoretical investigation on the ground state properties of a two dimensional ultra-cold Fermi superfluid with an anisotropic spin-orbit coupling (SOC). In the absence of Zeeman field, the system evolves from weak coupling BCS regime to strongly interacting BEC regime (BCS-BEC crossover) with increasing either the two-particle interaction strength or SOC parameters. We focused on the behaviors of pairing parameter and density of states (DOS) when increasing the anisotropic parameter of the SOC. Surprisingly, we discovered that the gap parameter decreases with increasing the anisotropic parameters, but the DOS at the Fermi surface shows non-monotonic behavior as a function of the anisotropic parameter. In the presence of the Zeeman field, we discussed a particular type of topological phase transition by obtaining the analytical result of the topological invariant and directly related this quantum phase transition with a sudden change of the ground state wave-function. Effects of higher partial wave pairing terms on this topological phase transition were briefly discussed.
\end{abstract}


\end{frontmatter}



\section{Introduction.}

In recent years, spin-orbit coupling effects (SOC) in condensed matter systems have received lots of interest \cite{winkler}. Firstly,
SOC is a key ingredient in realizing nontrivial topological phases \cite{volovic, nayak, qi, taylor, sarma}. For example, combined effects of SOC and an external Zeeman field in superconductor systems
can generate a non-Abelian quantum order \cite{sarma}.  Secondly, SOC can induce a nontrivial
spin-triplet pairing field which significantly changes
the properties of non-central-symmetric superconductors \cite{bergeret}. Furthermore, effects of SOC on the unconventional superconductivity also attract lots of attention recently \cite{luyang}.

In order to observe these novel phenomena, much effort has been invested to synthesize solid state materials with sizable
SOC. Another promising platform is the artificial materials, especially ultra-cold atoms system where SOC, Zeeman field can be readily generated and superfluidity has been observed with current experimental
technique \cite{spielman, wang, cheuk}. In ultra-cold atom community, construction of model systems on the Hamiltonian
level is now available \cite{dalibard, goldman}. Due to its highly controllability, ultra-cold atom system has been proven to be a ideal platform for the investigation of many fundamental problem in solid state chemistry and physics, such as the creation and manipulation of various crystalline structure using optical lattice trap and characterization of its energy band structure and other physical properties \cite{bloch}.

There are mainly two types of SOC, namely Rashba \cite{rashba} and Dresselhaus \cite{dress} SOC. In ultra-cold atoms systems, current experimental set-up can produce SOC with arbitrary
combination of these two types
of SOC \cite{socnp, socsc}. Many theoretical investigations have been performed to study effects of SOC on various superfluid properties \cite{ye, sala1, sala2, lee, levin, alberto, zhai1, pu1, liu, he1, jia, zinner, shenoy, iskin1, hu1, li, he2, zhai2, yu, liao, zlatko, zhai3, zhou1}. In the absence of Zeeman field, SOC can
produce a novel bound-state called Rashbons and
therefore induce a crossover from weakly correlated BCS to
strongly interacting BEC regime (BCS-BEC) even for very
weak particle-particle interaction \cite{shenoy}. Effects of anisotropic SOC on the ground state properties have been discussed in \cite{sala}. It was found that Rashba SOC is the optimal one for superconductivity/superfluidity. Anisotropy of SOC suppresses pairing and condensation.
 Furthermore, combined effect of SOC and Zeeman field can host a non-trivial topological order \cite{sarma, melo1, pu2, hu2, hu3, melo2, zhou2, iskin2, he3, chuanwei1, chuanwei2}. In two dimensional (2D) superfluid
 system with Rashba SOC, transition from
trivial superfluid state to non-trivial topological state can be characterized by a topological invariant which has been obtained analytically in \cite{zhou2}. The information contained in the topological invariant
and its physical consequence has also been discussed in \cite{zhou2}. However extension of this discussion to anisotropic SOC is still remained undone which is our main focus.

In this paper, we investigate the effect of the anisotropic SOC on the ground state properties of a 2D superfluid system using mean-field theory. In the
absence of Zeeman field, we calculate the density of states (DOS) at the Fermi surface with the self-consistent solutions of the mean-field number and gap equations.
Surprisingly, we find that the DOS at the fermi surface as a function of the anisotropic parameter is not monotonic and has local maximum for certain parameter space.
But the gap parameter decreases with increasing degree of anisotropy of SOC.
This means that gap parameter is not sensitive to the density of state at the Fermi surface and increasing DOS at the Fermi surface does not necessarily enhance pairing and the transition temperature. Furthermore, the maximum of the DOS as a function of the anisotropic parameter increases with increasing total strength of the SOC. In the presence of an external Zeeman field, we also study the topological phase transition characterized by a topological invariant. We obtain the analytical result of the topological invariant for arbitrary SOC and found that the anisotropic SOC does not change nature of the topological phase transition.

\section{Formalism.}

We consider an anisotropic SOC which can be written as an arbitrary
combination of Rashba and Dresselhaus type SOC. In momentum space, it can be
described by:

\begin{equation}
H_{soc}=\lambda _{R}\left( \sigma _{x}p_{y}-\sigma _{y}p_{x}\right) +\lambda
_{D}\left( \sigma _{x}p_{y}+\sigma _{y}p_{x}\right)   \label{soc}
\end{equation}
where $\lambda _{R}$ and $\lambda _{D}$ denote the Rashba and Dresselhaus SOC
parameters respectively and $\sigma _{i=x,y,z}$ are the Pauli matrices. The
system under consideration can be described by the Hamiltonian:
\begin{equation}
H=\int d^{2}\mathbf{r}\psi ^{\dagger }\left( \mathbf{r}\right) \left[
\varepsilon _{\mathbf{\hat{p}}}-h\sigma _{z}+H_{soc}\right] \psi \left(
\mathbf{r}\right) -g\int d^{2}\mathbf{r}\varphi _{\uparrow }^{\dagger
}\left( \mathbf{r}\right) \varphi _{\downarrow }^{\dagger }\left( \mathbf{r}%
\right) \varphi _{\downarrow }\left( \mathbf{r}\right) \varphi _{\uparrow
}\left( \mathbf{r}\right)
\end{equation}
where $g>0$ is the contact interaction parameter and $\varphi _{\sigma
\left( =\uparrow ,\downarrow \right) }\left( \mathbf{r}\right) $ and $%
\varphi _{\sigma }^{\dagger }\left( \mathbf{r}\right) $ are the annihilation
and creation field operators, respectively,  $\psi \left( \mathbf{r}\right) =%
\left[ \varphi _{\uparrow }\left( \mathbf{r}\right) ,\varphi _{\downarrow
}\left( \mathbf{r}\right) \right] ^{T}$ and kinetic energy $\varepsilon _{%
\mathbf{\hat{p}}}=\mathbf{\hat{p}}^{2}/2m-\mu $ with $m$, $\mu $\ and $h$
being the mass of the Fermi atoms, the chemical potential and the effective
Zeeman field, respectively. For simplicity we set $\hbar =1$ throughout this
paper. As can be seen from Eq. ({\ref{soc}}), the system is isotropic when $%
\lambda _{D}=0$ or $\lambda _{R}=0$ and anisotropic when $\lambda
_{D}=\lambda _{R}$. For convenience, we define an anisotropic parameter as
\begin{equation}
\eta =\frac{\lambda _{D}}{\lambda _{R}}
\end{equation}
When $\eta $ increases from $0$ to $1$, the system evolves from isotropic
Rashba case to anisotropic case with equal Rashba and Dresselhaus SOC.

Within mean-field theory, the interacting part can be approximated by $-\int
d^{2}\mathbf{r}\left( \Delta \left( \mathbf{r}\right) \varphi _{\uparrow
}^{\dagger }\left( \mathbf{r}\right) \varphi _{\downarrow }^{\dagger }\left(
\mathbf{r}\right) +h.c.\right) +\int d^{2}\mathbf{r}\left\vert \Delta \left(
\mathbf{r}\right) \right\vert ^{2}/g$ with $\Delta \left( \mathbf{r}\right) $
being the pairing field. For our purpose, we only consider translational
invariant solutions where the paring field becomes a constant $\Delta \left(
\mathbf{r}\right) =\Delta $. Therefore, the Hamiltonian can be represented in momentum space and its matrix form reads: $H=\sum_{\mathbf{p}>0}\Phi _{\mathbf{p}}^{\dagger
}H_{BdG}\left( \mathbf{p}\right) \Phi _{\mathbf{p}}+\sum_{\mathbf{p}%
}\varepsilon _{\mathbf{p}}+V\Delta ^{2}/g$ where $V$ denotes the size of the
system, $\Phi _{\mathbf{p}}=\left[ a_{\mathbf{p},\uparrow },a_{\mathbf{p}%
,\downarrow },a_{-\mathbf{p},\uparrow }^{\dagger },a_{-\mathbf{p},\downarrow
}^{\dagger }\right] ^{T}$ and the BdG Hamiltonian $H_{BdG}\left( \mathbf{p}%
\right) $ is
\begin{equation}
H_{BdG}\left( \mathbf{p}\right) =\left[
\begin{array}{cccc}
\varepsilon _{\mathbf{p}}-h & \Gamma _{\mathbf{p}} & 0 & -\Delta  \\
\Gamma _{\mathbf{p}}^{\ast } & \varepsilon _{\mathbf{p}}+h & \Delta  & 0 \\
0 & \Delta  & -\varepsilon _{\mathbf{p}}+h & \Gamma _{\mathbf{p}}^{\ast } \\
-\Delta  & 0 & \Gamma _{\mathbf{p}} & -\varepsilon _{\mathbf{p}}-h%
\end{array}%
\right]   \label{ham}
\end{equation}%
with $\Gamma _{\mathbf{p}}=\lambda _{R}\left( p_{y}+ip_{x}\right) +\lambda
_{D}\left( p_{y}-ip_{x}\right) $.

Using the standard diagonalization procedure, we obtain the ground-state
free energy $E_{g}=\sum_{\mathbf{p},s\mathbf{=\pm }}\left( \varepsilon _{%
\mathbf{p}}-E_{\mathbf{p},s}\right) /2+V\Delta ^{2}/g$ where the excitation
spectrum $E_{\mathbf{p},s}=\sqrt{\mathcal{E}_{\mathbf{p},s}^{2}+\Delta _{%
\mathbf{p},2}^{2}}$ with $\mathcal{E}_{\mathbf{p},s}=E_{\mathbf{p}}-s\sqrt{%
h^{2}+\left\vert \Gamma _{\mathbf{p}}\right\vert ^{2}}$, $E_{\mathbf{p}}=%
\sqrt{\varepsilon _{\mathbf{p}}^{2}+\Delta _{\mathbf{\ p},1}^{2}}$, $\Delta
_{\mathbf{\ p},1}=\Delta \left\vert \cos \theta _{\mathbf{p}}\right\vert $, $%
\Delta _{\mathbf{p},2}=\Delta \sin \theta _{\mathbf{p}}$ and $\theta _{%
\mathbf{p}}=\pi -\arctan \left( \left\vert \Gamma _{\mathbf{p}}\right\vert
/h\right) $. From variation of ground state energy with respect to the gap
parameter and chemical potential, we obtain the gap and number equations
\begin{eqnarray}
\frac{1}{g} &=&\frac{1}{V}\sum_{\mathbf{p},s}\frac{1+s\cos \theta _{\mathbf{p%
}}\frac{h}{E_{\mathbf{p}}}}{4E_{\mathbf{p},s}},  \label{gap} \\
N &=&\frac{1}{2}\sum_{\mathbf{p},s}\left( 1-\frac{\mathcal{E}_{\mathbf{p},s}%
}{E_{\mathbf{p},s}}\frac{\varepsilon _{\mathbf{p}}}{E_{\mathbf{p}}}\right) .
\label{number}
\end{eqnarray}
Divergence of the integral over momenta in Eq. ({\ref{gap}}) is removed by
replacing contact interaction parameter $g$ by binding energy $E_{b}$
through $V/g=\sum_{\mathbf{p}}1/\left( 2\epsilon _{\mathbf{p}}+E_{b}\right) $%
.

Furthermore, the ground-state wave-function can be obtained as:

\begin{equation}
\left\vert G\right\rangle =\prod\limits_{\mathbf{p}>0,s}\left( u_{\mathbf{p}%
,s}+e^{is\varphi _{\mathbf{p}}}v_{\mathbf{p},s}\beta _{\mathbf{p}%
,s}^{\dagger }\beta _{-\mathbf{p},s}^{\dagger }\right) \left\vert
g\right\rangle   \label{wavefunction}
\end{equation}

where $\beta _{\mathbf{p},s}=u_{\mathbf{p}}c_{\mathbf{p},s}-v_{\mathbf{p}%
}c_{-\mathbf{p},-s}^{\dagger }$ with $u_{\mathbf{p}}=\sqrt{\left(
1+\varepsilon _{\mathbf{p}}/E_{\mathbf{p}}\right) /2}$ and $u_{\mathbf{p}%
}^{2}+v_{\mathbf{p}}^{2}=1$, $c_{\mathbf{p},s}=\sin \left( \theta _{\mathbf{p%
}}/2\right) a_{\mathbf{p},s}-s\cos \left( \theta _{\mathbf{p}}/2\right)
e^{is\varphi _{\mathbf{p}}}a_{\mathbf{p},-s}$ with $\varphi _{\mathbf{p}%
}=\arctan \left[ \left( \lambda _{R}-\lambda _{D}\right) p_{x}/\left(
\lambda _{R}+\lambda _{D}\right) p_{y}\right] $ and
\begin{equation}
\left[
\begin{array}{c}
u_{\mathbf{p},s} \\
v_{\mathbf{p},s}%
\end{array}%
\right] =\sqrt{\frac{1}{2}\left( 1\pm \frac{\mathcal{E}_{\mathbf{p},s}}{E_{%
\mathbf{p},s}}\right) }.  \label{uvs}
\end{equation}

\section{Balanced case.}

In the absence of Zeeman field,  $h=0$, the ground
state properties have been investigated in \cite{sala}. The self-consistent solution
of the gap and number equations show that the pairing parameter $\Delta $
decreases with increasing anisotropic parameter $\eta $. In this paper, we
focus on the dependence of pairing parameters on the DOS at the
Fermi surface. For $h=0$, Hamiltonian in the helicity basis $c_{\mathbf{p},s}
$ becomes

\begin{equation}
H=\frac{\Delta ^{2}}{g}+\sum_{\mathbf{p,}s\mathbf{=\pm }}\mathcal{E}_{%
\mathbf{p},s}c_{ks}^{\dagger }c_{ks}-\frac{\Delta }{2}\sum_{k}\left(
e^{i\varphi _{k}}c_{k,+}^{\dagger }c_{-k,+}^{\dagger }+e^{-i\varphi
_{k}}c_{k,-}^{\dagger }c_{-k,-}^{\dagger }\right)
\end{equation}

As can been seen from the above equation, pairing happens only between the
same helicity basis. And DOS of the helicity basis is defined as

\begin{equation}
\rho _{F}=\sum_{\mathbf{p,}s\mathbf{=\pm }}\delta \left( -\mathcal{E}_{%
\mathbf{p},s}\right)
\end{equation}

\begin{figure}[tbh]
\includegraphics[width=\columnwidth,height=60mm]{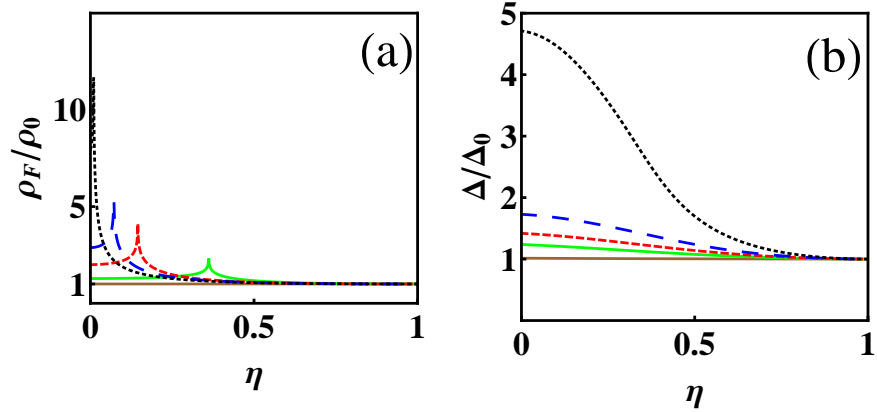}
\caption{(Color online) DOS at the Fermi surface $\rho _{F}$ (a) and pairing parameter $\Delta$ (b) as functions of the anisotropic parameter $\eta$. Different lines correspond to different values of $\tilde{\lambda}$. The brown solid, green solid, red dashed, long blue dashed and black dotted lines correspond to $\tilde{\lambda} = 0.2$, $0.8$, $1.0$, $1.2$ and $2.0$ respectively.}
\label{gapdos1}
\end{figure}

\begin{figure}[tbh]
\includegraphics[width=\columnwidth,height=60mm]{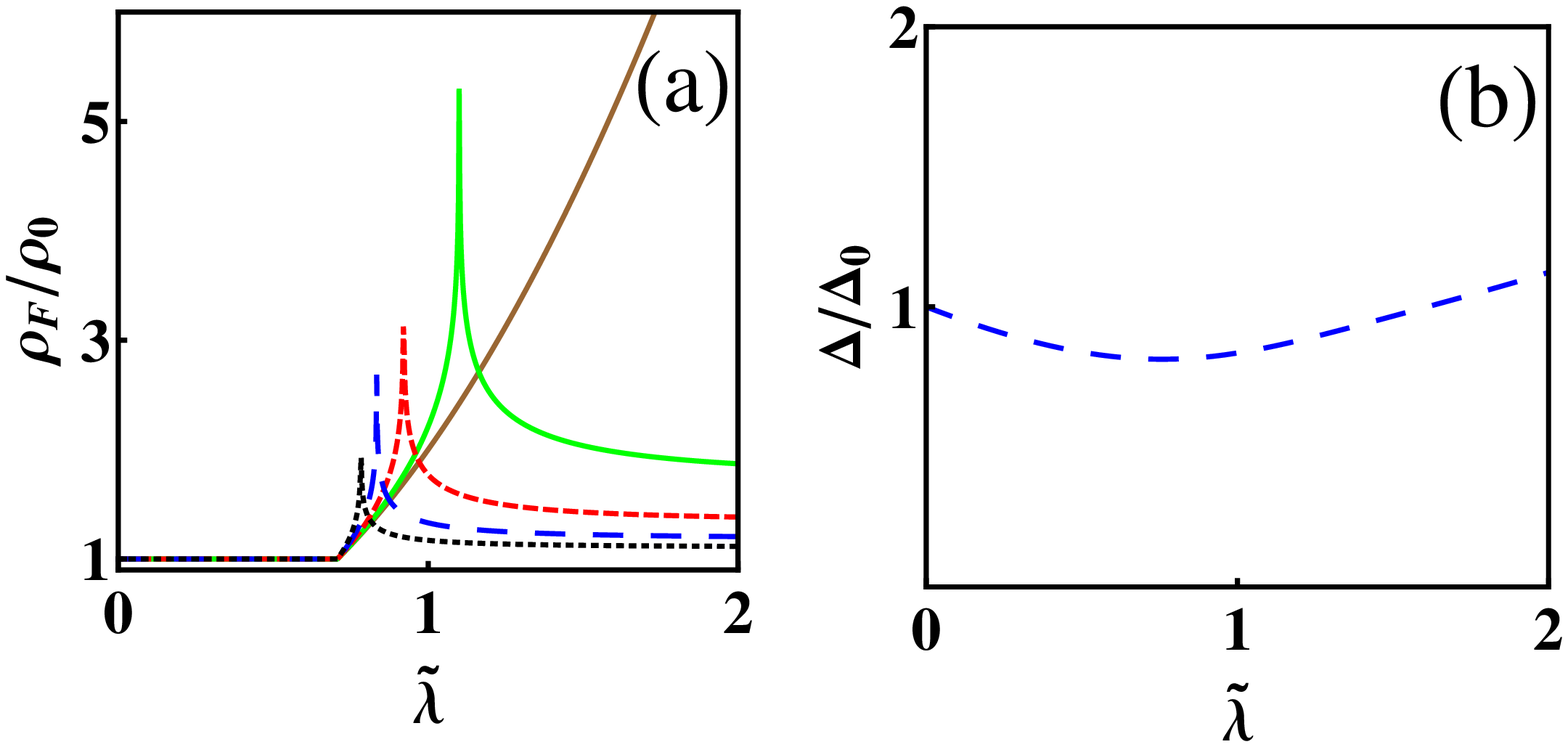}
\caption{(Color online) DOS at the Fermi surface $\rho _{F}$ (a) and pairing parameter $\Delta$ (b) as functions of $\tilde{\lambda}$. Different lines correspond to different values of $\eta$. In (a), the brown solid, green solid, red dashed, long blue dashed and black dotted lines correspond to $\eta= 0$, $0.1$, $0.2$, $0.3$ and $0.4$ respectively.In (b), we set $\eta=0.2$ without loss of generality.}
\label{gapdos2}
\end{figure}

Performing the momentum integral, we obtain

\begin{equation}
\rho _{F}=\frac{\Theta \left( \mu \right) }{2\pi }+\frac{\Theta \left( -\mu
\right) }{\left( 2\pi \right) ^{2}}\int_{0}^{2\pi }dx\frac{\Psi \left(
x\right) \Theta \left[ \Psi ^{2}\left( x\right) +\mu \right] }{\sqrt{\Psi
^{2}\left( x\right) +\mu }}  \label{dos}
\end{equation}
with $\Psi \left( x\right) =\sqrt{\lambda _{R}^{2}+\lambda _{D}^{2}-2\lambda
_{R}\lambda _{D}\cos \left( 2x\right) }$. As already known that, in the absence of SOC, $\Delta $
depends on $\rho _{F}$ explicitly in the weak interacting limit because
pairing happens only around the Fermi surface. Meanwhile, for an
isotropic Rashba SOC,  $\Delta $ and $\rho _{F}$ both increase when
increasing the SOC strength \cite{zhai3}. Therefore, it is believed that $\Delta $
depends on $\rho _{F}$ in a monotonic manner. And increasing density of
states at the Fermi surface is considered as an efficient way of increasing
the pairing strength and transition temperature. However, we find that the
anisotropic nature of the SOC significantly changes this picture. The\ numerical results
of DOS and gap parameter are presented in Fig. \ref{gapdos1} and Fig. \ref{gapdos2}. Without loss of generality,
in the numerical calculations, we have set $E_{b}=0.5E_{F}$ with $%
E_{F}=k_{F}^{2}/2m$ and $k_{F}=\sqrt{2\pi n}$. Fig. \ref{gapdos1} represents DOS and gap parameter as functions of anisotropic parameter and different lines correspond to
different $\lambda =\sqrt{\lambda _{R}^{2}+\lambda _{D}^{2}}$, $\rho _{0}$
and $\Delta _{0}$ are the DOS at the Fermi surface and gap parameter for $\lambda =0$. Fig. \ref{gapdos2} shows DOS and gap parameter as functions of dimensionless parameter $\tilde{\lambda}=m\lambda/k_F$  and different lines correspond to
different $\eta$.

From Fig. \ref{gapdos1}(a), we find that the DOS at the Fermi surface as a function of
the anisotropic parameter is not a monotonic function. However, as seen from Fig. \ref{gapdos1} (b), the gap
parameter $\Delta $ decreases as $\eta $ increases and $\Delta $ reduces to $%
\Delta _{0}$ for equal Rashba and Dresselhaus SOC. Therefore, pairing does
not increase monotonically with DOS at the Fermi surface. Furthermore, for
large enough $\lambda $, it has a maximum value and for small $\lambda $,
the chemical potential remains positive and $\rho _{F}=\rho _{0}$ as can be
seen from Eq. ({\ref{dos}}
) and the brown solid line in Fig. \ref{gapdos1} (a). Last but not least, the
maximum value of DOS increases with increasing $\lambda $.

More interestingly, as can be seen from Fig. \ref{gapdos2}, both the DOS at the Fermi surface and gap are non-monotonic functions of $\lambda $ for for $0<\eta<1$. When $\eta=0$, the system is isotropic and $\Delta $ increases with increasing $\lambda$ \cite{he2,zhai3,zhou1}. However, for $\eta=1$, the SOC terms reduces to equal Rashba and Dresselhaus case. In this case and without Zeeman field, the SOC does not affect the thermodynamic properties and therefore $\Delta $ does not change with increasing $\lambda$. Furthermore, for $0<\eta<1$, the gap parameter as a function of $lambda$ has a local minimum as can be seen from Fig. \ref{gapdos2} (b).

\section{Imbalanced case.}
In the presence of an external Zeeman field, the
ground state of the system under consideration becomes far more complex. Many
exotic phases may appear and the ground state phase diagram has been
investigated extensively \cite{melo1, pu2, hu2, hu3, melo2, zhou2, iskin2, he3, chuanwei1, chuanwei2}.
Most interestingly, there is a topological phase transition
driven by Zeeman field which is our main focus. The critical Zeeman
field reads $h_{c}=\sqrt{\mu ^{2}+\Delta ^{2}}$. For $h<h_{c}$, the system
is in a trivial gapped superfluid state. When $h>h_{c}$, the ground state is topologically
nontrivial and is characterized by a nonzero topological invariant $\mathcal{%
N}$ which is defined as \cite{sarma, volovic} $\mathcal{N}=1/2\pi
\int_{-\infty }^{+\infty }d^{2}\mathbf{p}B\left( \mathbf{\ p}\right) $ with
the Berry curvature being given by
\begin{equation}
B\left( \mathbf{p}\right) =-i\sum_{E_{\mathbf{p}}^{\alpha }<0}\left[
\partial _{p_{x}}\mathbf{u}_{\mathbf{p},\alpha }^{\dagger }\partial _{p_{y}}%
\mathbf{u}_{\mathbf{p},\alpha }-\partial _{p_{y}}\mathbf{u}_{\mathbf{p}%
,\alpha }^{\dagger }\partial _{p_{x}}\mathbf{u}_{\mathbf{p},\alpha }\right]
\label{bc}
\end{equation}%
where $\mathbf{u}_{\mathbf{p},\alpha =1,2,3,4}$ are the eigenvectors of Eq.
( {\ref{ham}}) corresponding to the eigenvalues $-E_{\mathbf{p},+},E_{%
\mathbf{p},+},-E_{\mathbf{p},-},E_{\mathbf{p},-}$, respectively. Following
the same procedure in \cite{zhou2}, we obtain the eigen states as $\mathbf{u}_{%
\mathbf{p},s=\pm }=\left[ e^{is\varphi _{\mathbf{p}}}F_{\mathbf{p},s}^{1},F_{%
\mathbf{p},s}^{2},F_{\mathbf{p},s}^{3},e^{is\varphi _{\mathbf{p}}}F_{\mathbf{%
p},s}^{4}\right] ^{T}$ with
\begin{eqnarray}
F_{\mathbf{p},s}^{1} &=&u_{\mathbf{p}}\sin \frac{\theta _{\mathbf{p}}}{2}v_{%
\mathbf{p},s}-v_{\mathbf{p}}\cos \frac{\theta _{\mathbf{p}}}{2}u_{\mathbf{p}%
,s} \\
F_{\mathbf{p},s}^{2} &=&u_{\mathbf{p}}\cos \frac{\theta _{\mathbf{p}}}{2}v_{%
\mathbf{p},s}+v_{\mathbf{p}}\sin \frac{\theta _{\mathbf{p}}}{2}u_{\mathbf{p}%
,s} \\
F_{\mathbf{p},s}^{3} &=&u_{\mathbf{p}}\sin \frac{\theta _{\mathbf{p}}}{2}u_{%
\mathbf{p},s}+v_{\mathbf{p}}\cos \frac{\theta _{\mathbf{p}}}{2}v_{\mathbf{p}%
,s} \\
F_{\mathbf{p},s}^{4} &=&u_{\mathbf{p}}\cos \frac{\theta _{\mathbf{p}}}{2}u_{%
\mathbf{p},s}-v_{\mathbf{p}}\sin \frac{\theta _{\mathbf{p}}}{2}v_{\mathbf{p}%
,s}.
\end{eqnarray}

The only difference here is the anisotropic SOC characterized by the phase
factor $\varphi _{\mathbf{p}}=\arctan \left[ \vartheta p_{x}/p_{y}\right] $
with $\vartheta =\left( \lambda _{R}-\lambda _{D}\right) /\left( \lambda
_{R}+\lambda _{D}\right) $. Simple algebraic manipulation leads to
\begin{equation}
B\left( \mathbf{p}\right) =\partial _{p_{y}}\varphi _{\mathbf{p}}\partial
_{p_{x}}F_{\mathbf{p}}-\partial _{p_{x}}\varphi _{\mathbf{p}}\partial
_{p_{y}}F_{\mathbf{p}}=-\frac{\vartheta \mathbf{p}}{p_{y}^{2}+\left(
\vartheta p_{x}\right) ^{2}}\cdot \mathbf{\nabla }F_{\mathbf{p}}
\label{berry2D}
\end{equation}%
and $F_{\mathbf{p}}=\sum_{\alpha =1,4,s}s\left( F_{\mathbf{p},s}^{\alpha
}\right) ^{2}$. Clearly the Berry curvature $B\left( \mathbf{p}\right) $ is
anisotropic. However, by proper scaling of the integral variables, the
topological invariant does not depend on the anisotropic parameter and we
obtain
\begin{equation}
\mathcal{N}=F_{\mathbf{0}}=v_{\mathbf{0},+}^{2}=\theta \left( h-h_{c}\right)
.  \label{invariant}
\end{equation}

From this we can see that the topological phase transition corresponds to a
sudden change of the ground state wave function at zero momentum
characterized by $v_{\mathbf{0,}+}^{2}$. Consequently, there is a sudden
change of the ground-state wave-function associated with the component of
triplet pairing of the quasi-particles denoted by $\beta _{\mathbf{p},+}$ at
zero momentum. This is also reflected in the momentum distribution as can be
seen from Eq. (\ref{number}) that $\mathcal{E}_{\mathbf{0},+}/E_{\mathbf{0}%
,+}=sign(h_{c}-h)$. This unique property provides a conclusive evidence that the topological phase
transition can be determined by measuring the momentum distribution in cold
atom experiments.

Furthermore,
in the presence of higher partial wave pairing terms, taking $p$ and $d$
wave pairing symmetry for example, since the topological phase transition
depends only on the zero momentum parts of the ground state wave function,
the $p$ wave pairing does not affect the topological phase transition while $d$
wave does \cite{sarma}.

\section{Conclusion.}

We investigated the ground-state properties of a 2D
Fermi superfluid system in the presence of a general anisotropic SOC and Zeeman coupling that
supports non-trivial topological order.\ Particularly, we found that
increasing the DOS at the Fermi surface is not a sufficient way of
obtaining large $\bigtriangleup $ and high transition temperature. For the topological phase transition driven by an external Zeeman field, we found that the anisotropic nature of the
system induced by an anisotropic SOC does not change the topological phase
transition. And from the analytical result of the topological invariant, we discovered that the topological phase transition can be determined by
measuring the momentum distribution in cold atomic experiments.

\section{Acknowledgements.}
This work has been supported by the National
Natural Science Foundation of China under Grant 51331006, 51590883, 11204321, the National Basic Research Program (No.2017YFA0206302) of China and the project of Chinese academy of Sciences with grant number KJZD-EW-M05-3.

\end{document}